# Techno Economic Modeling for Agrivoltaics: Can Agrivoltaics be more profitable than Ground mounted PV?

Habeel Alam, Muhammad Ashraful Alam, *Fellow, IEEE* and Nauman Zafar Butt, *Member, IEEE*

*Abstract—* Agrivoltaics (*AV*) is a dual land-use approach to collocate solar energy generation with agriculture for preserving the terrestrial ecosystem and enabling food-energy-water synergies. Here, we present a systematic approach to model the economic performance of *AV* relative to standalone ground-mounted PV (*GMPV*) and explore how the module design configuration can affect the dual food-energy economic performance. A remarkably simple criterion for economic feasibility is quantified that relates the land preservation cost to dual food-energy profit. We explore case studies including both high and low value crops under fixed tilt bifacial modules oriented either along the conventional North/South (*N/S*) facings or vertical East/West (*E/W*) facings. For each module configuration, the array density is varied to explore an economically feasible design space relative to *GMPV* for a range of module to land cost ratio ($M_L$) – a location-specific indicator relating the module technology (hardware and installation) costs to the soft (land acquisition, tax, overheads, *etc.*) costs. To offset a typically higher *AV* module cost needed to preserve the cropland, both *E/W* and *N/S* orientated modules favor high value crops, reduced (<60%) module density, and higher $M_L$ (> 25). In contrast, higher module density and an increased feed-in-tariff (*FIT*) relative to *GMPV* are desirable at lower $M_L$. The economic trends vary sharply for $M_L < 10$ but tend to saturate for $M_L > 20$. For low value crops, ~15% additional *FIT* can enable economic equivalence to *GMPV* at standard module density. The proposed modeling framework can provide a valuable tool for *AV* stakeholders to assess, predict, and optimize the techno-economic design for *AV*.

## I. INTRODUCTION

The global agriculture requires a projected capacity to feed around 10 billion people by 2050 [1]. The preservation of agricultural lands, sustainable increase in crops' yield and protection against the climate change are some of the primary approaches to meet this global challenge. Similar to the growing food needs, the global energy consumption is predicted to increase by nearly 50% over the next 30 years [2] which necessitates an enormous growth in the renewable energy generation, including solar and wind, to save the climate. Sustainable food-energy solutions also require an effective use of the land to preserve the terrestrial ecosystems and biodiversity. The conventional ground mounted photovoltaic (*GMPV*) systems are not designed for the utilization of their covered arable land area for dual food-energy production. As the global *GMPV* installations are increasing rapidly, concerns over the land use change, food security, and biodiversity preservation are continuously on the rise [3-6]. Moreover, with falling prices for solar power technology and increasing pressure of climate vulnerabilities, famers in many countries are tempted to convert their agricultural land into *GMPV* [7].

An innovative approach to address these issues are the dual land usage systems called agrivoltaics (*AV*) which have recently gained a widespread popularity [8, 9]. An *AV* system utilizes the same land for the dual production of crops and energy by elevating the panels above ground and configuring them to facilitate agricultural operations [10]. The concept of *AV* was initially proposed by Goetzberger and Zastrov back in 1981[11]. During the last decade, many academic and commercial scale *AV* installations [12-16] have been reported that indicate attractive synergies for the food-energy-water nexus including a higher water use efficiency, increased yield for the selected crops, and a cooling effect for solar module resulting in higher energy yield. In many countries, government policies also support *AV* and currently ~2000 *AV* systems with cumulative capacity of 2.8$GW_p$ have been installed across the globe [12-14]. Many field studies and modeling work have also been reported for assessing and predicting the performance for different *AV* module configurations and crops [17-22]. Some of the common *AV* module configurations are shown in Fig 1 with a comparison with typical *GMPV*.

While a win-win situation for food-energy attracts a lot of interest for *AV*, its economic feasibility for all the stakeholders including *PV* investors, farmers, and policymakers is critically important for its widespread acceptability. Although significant research has been reported on *AV* module technologies, crop-specific field experiments, and food-energy yield modeling [12-23], economic aspects for *AV* are relatively less understood. In particular, the economic tradeoffs as a function of various module configurations, land-specific costs, energy tariffs, and crop profits have not been modeled in comparison with *GMPV*. The installation of *AV* modules is typically more expensive as compared to *GMPV* due to an elevated mounting and customized foundations that are typically needed to facilitate agricultural operations on the same land [24]. A higher levelized cost of electricity (*LCOE*) for *AV* as compared to *GMPV* is therefore considered to be the land preservation cost for the *AV* system [12]. A recent study by NREL [25] estimates the land preservation costs for *AV* to be 50% to 20% of the premium costs of the *GMPV* for the *N/S* faced fixed tilt and the vertical $E/W$ faced systems, respectively.

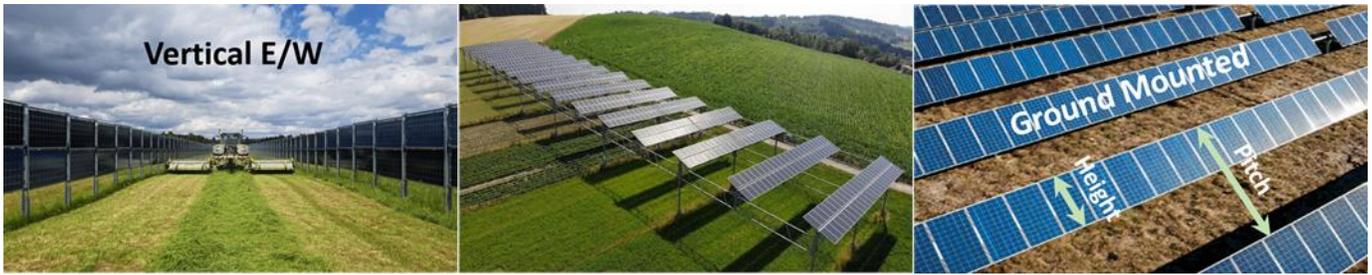

Figure 1. a) Vertical East-West (E/W) faced bifacial agrivoltaics (AV) farm [26], b) North south (N/S) faced agrivoltaics (AV) farm (source: BayWa r.e.) [12]. c) A typical ground mounted photovoltaic (GMPV) system with pitch and height labelled.

Despite a higher $LCOE$, an additional revenue from crops can make $AV$ economically superior to $GMPV$ provided it could offset the land preservation cost. A recent field study [12] compares the economic performance of a 194.4 $KW_p$ $AV$ system which grows potatoes and winter wheat with $GMPV$. A simple model that is based on price-performance ratio is used to evaluate the economic performance, where the price is the land preservation cost and the performance benefit comes from the crop revenue. The $AV$ modules which are elevated 5m above the ground results in 38% higher $LCOE$ in comparison with $GMPV$. The revenue from potatoes is shown to be high enough to offset the increase in $LCOE$, making $AV$ more profitable than $GMPV$ although the biomass yield for potatoes drops to ~87% for $AV$ as compared to the full sun condition. The revenue from the winter wheat is however significantly lower which could not compensate for the land preservation cost for $AV$. Although this modeling approach is useful, the focus of the study is limited to the given field experiment and the effect of varying the module design, land costs, and $FIT$ are not explored.

In another recent study [27], an analytical framework determining the economic benefits and adoption potential of $AV$ has been proposed. The economic performance for $AV$ is evaluated from the perspective of maintaining the farmer's profitability with respect to the conventional agriculture farm. While the model applies well to the farmer's economic perspective, it does not explore the solar investor's profitability, in particular, the impact of land preservation cost on the solar energy profit and the economic performance of $AV$ relative to the conventional $GMPV$. The economic performance for $AV$ relative to $GMPV$ and roof-top $PV$ is explored in [24] under a set of economic assumptions. The study reports that $GMPV$ systems are about 33% cheaper than $AV$ due to halved costs for installation, balance of plant and support infrastructure but the net present value for $AV$ could generate more profit towards the end of the project lifetime. The potential of rainfed $AV$ for ground water stressed irrigated regions across the world is studied in [28] by integrating solar, crop, hydrology and financial models. A simulation-based study reported in [29] shows the advantage of $AV$ over the traditional farming systems which rely on diesel engine for irrigation. The efficient usage of land is also demonstrated by calculation of land equivalent ratio.

Even though the reported economic studies on $AV$ demonstrate important trends and case studies, there is a lack of a systemic techno-economic evaluation that could quantify the relationship between the module array configurations and the food-energy economic performance. A holistic model is needed to explore the effect of varying $AV$ module configurations under a broad range of economic factors including the soft land costs which can widely vary on local as well as global scales and the module related system costs which have a lesser location dependence but can strongly vary as a function of the module technology and system configuration. Moreover, in contrast to the conventional food-energy systems, where standard parameters such as $LCOE$ and farm profits are routinely used, useful metrics to quantify the combined food-energy economic performance are lacking for $AV$ which makes the system design optimization difficult.

In this paper, we present a techno-economic model based on convenient metrics to quantitatively explore the relationship of module array configurations with the economic performance for $AV$ as a function of technology-specific land preservation costs, crop income, energy tariffs, and land-specific soft land costs. For the first time, this study presents a systematic approach to explore: (i) how the module array density for $AV$ affects its overall economic performance? (ii) what profit margins (crop rotations) could offset the land preservation cost? (iii) what is the economic threshold to invest on $AV$ infrastructure to preserve an agriculture land for a given crop and location? (iv) what is the minimum adjustment in $FIT$ if a given $AV$ system needs to meet the economic equivalence to $GMPV$? (v) what is an optimal design space for the module arrays in terms of spatial density to maximize the economic performance? (vi) how does the economic performance vary with the location specific costs including land acquisition, overheads, and taxes? and (vii) how does the $AV$ economics vary as the module orientation is varied from the traditional $N/S$ faced tilted to vertically tilted $E/W$ faced configuration?

The rest of the paper is arranged as follows: In section II, we report the methodology and mathematical modelling of this techno-economic framework highlighting its major assumptions and components. In Section III, we apply the framework to assess the economic feasibility of $AV$ for two different orientations ($N/S$ and vertical bifacial $E/W$) across two simulated crop rotations for Khanewal located in Southern Punjab, Pakistan. Results obtained after application of framework are also discussed in section III. Section IV reports conclusion and limitations (including possible modification).

| | | | |
|---|---|---|---|
| **Nomenclature** | | $\kappa$ | Land preservation cost |
| | | $LCOE$ | Levelized cost of electricity |
| $AV$ | Agrivoltaics | $M_L$ | Module-to-land cost ratio |
| $A_L$ | Land area | $P_{AV}$ | Annual energy profit from AV |
| $A_{L,AV}$ | AV land area | $P_C$ | Annual crop profit in AV ($/year) |
| $A_{L,PV}$ | GMPV land area | $P_{C\,open}$ | Annual crop profit in open agriculture ($/year) |
| $A_M$ | Module area | $P_{GMPV}$ | Annual energy profit from AV |
| $c_L$ | Cost (per unit land area) related to land | $P'_C$ | Normalized crop profit |
| $c'_L$ | Normalized land related costs | $\rho$ | Overall food-energy profit for AV |
| $c_M$ | Cost (per unit module area) related to module technology | $p/h$ | Design parameter representing the module array density |
| $d$ | Depreciation rate | $r$ | Discount rate |
| $GW_P$ | Gigawatt-peak | $\psi$ | Normalized parameter quantifying the combined effect of costs related to modules and land |
| $ha$ | Hectare | $YY_T$ | Annual energy production |
| $kW_P$ | Kilowatt-peak | $Y_{PAR}$ | AV crop yield relative to open field |

## II. MATHATICAL MODELLING

### A. Basic Economic Model

While comparing the relative economic performance for $AV$ with $GMPV$, the basic premise of the model is that the land preservation cost needs to be offset by the net food-energy profit from the land. Additional constraints such as the minimum threshold yield for the crop production could further be applied. The extra costs incurred for $AV$ due to a customized mounting need to be offset by additional revenue generated from crops:

$$P_{AV} + P_c \geq P_{GMPV} \qquad (1)$$

where $P_{AV}, P_{GMPV}$ are the annual energy profit from $AV$ and $GMPV$, respectively, and $P_c$ denotes the profit from crops in $/year. Similarly, from the crop's perspective, any loss in the $AV$ crop yield relative to the open field condition needs to be balanced by the energy profit:

$$P_{AV} + P_c \geq P_{C_{open}} \qquad (2)$$

where $P_{C_{open}}$ denotes the profit from crops in open agriculture farm in $/year. An investment on $AV$ can be economically attractive when (1) and (2) are both satisfied. In most practical cases worldwide, the annual profit per unit land area from solar energy can be significantly greater than the annual profit per unit land area from agriculture. We therefore assume that (1) implies (2) and focus on exploring conditions that could satisfy (1). It is implicit that the overall profit of $AV$ (left hand side of (1) and (2)) is of interest in our approach rather than the individual energy and food profits while assuming that the yields do not drop below the limits imposed by local policy. Maximizing the overall food-energy profit is best applicable for the case when a single entity owns the revenues from $AV$ energy and the agriculture al production. If, however, multiple stakeholders own the food and energy profits, bilateral contracts, and the government policies could be defined to ensure the mutually agreed profitability of the individual entities in support of the land preservation. Future work is needed to explore various scenarios in this context.

We can express the energy profit in the form of feed-in-tariff ($FIT$), $LCOE$, and annual energy production ($YY_T$) in (1):

$$(FIT_{PV} - FIT_{AV} + LCOE_{AV} - LCOE_{PV}) \times YY_T \leq P_C \qquad (3)$$

where the subscripts $AV$ and $PV$ denotes the agrivoltaics and $GMPV$ systems, respectively. We assume an identical annual energy generation capacity for $AV$ and $GMPV$. Depending upon the $AV$ module configuration, this may result in different land and module areas for the $AV$ vs. $GMPV$. We will discuss the case of different $FIT$ for $AV$ and $GMPV$ in the next subsection. Here we assume that there is no difference in the $FIT$, so we can write:

$$(LCOE_{AV} - LCOE_{PV}) \times YY_T \leq P_C \qquad (4)$$

We can express $LCOE$ as [30]:
$$LCOE = \frac{c_M \cdot A_M + c_L \cdot A_L}{YY \cdot A_M \cdot \chi} = \frac{M_L + p/h}{YY \cdot \chi / c_L} \qquad (5)$$

where $c_M$ is cost (per unit module area) related to module technology (including balance of system), and $c_L$ is cost (per unit land area) related to land (including overhead, developer profit, and taxes), including capital and the operational costs (excluding the residual values) evaluated over the lifetime. $A_M$ and $A_L$ refer to module and land areas, respectively, $p/h$ is a design parameter representing the module array density where $p$ is the pitch (row to row distance) and $h$ is the vertical dimension of the module. $M_L = c_M/c_L$ is a ratio that depicts the effective costs related to the module technology to that related to the location specific costs for the land. $M_L$ can vary across locations and strongly depends on local policies. Typically reported values for US ranges between $10 - 20$ for $GMPV$ [12, 24, 30] Globally, reported values of $M_L$ for $GMPV$ ranges between 5-15, but due to the typically higher land requirement for $AV$, $M_L$ is expected to be higher for $AV$ in comparison with $GMPV$ [31]. $\chi \equiv \sum_{k=1}^{Y}(1-d)^k(1+r)^{-k}$, where $d$, and $r$ are rates for depreciation and discount rates, respectively. $c_M$ for a specific $GMPV$ technology does not change significantly across global locations (slight variations are possible due to the local policies related to taxation, import, material costs, and labor). $c_L$, on the other hand, can vary significantly across locations, both across the country and on global scales depending upon the type of the land, (e.g., urban vs. rural), soil fertility, and water availability, *etc*.

We can rewrite (4) by dividing $LCOE_{PV}$ on both sides and simplifying using (5):

$$\frac{\left(M_L+\frac{p}{h}\right)_{AV}}{\left(M_L+\frac{p}{h}\right)_{PV}} \leq \frac{P_C}{\left(M_L+\frac{p}{h}\right)_{PV} \times \left(\frac{c_L}{\chi}\right) \times A_M} + 1 \quad (6)$$

After simplifying (6), we get (see appendix):

$$\left(\frac{c_{M_{AV}}}{c_{M_{PV}}}\right) \equiv \kappa \leq \alpha \cdot P_C \cdot \chi - \left(\frac{1}{Y_{PV}} - \frac{A_{L,PV}}{A_{L,AV}}\right) \alpha \cdot c_L + Y_{PV} \quad (7)$$

where $\kappa$ is the ratio of module related costs for $AV$ relative to that for $GMPV$ and represents land preservation cost, $A_{L,PV}$ and $A_{L,AV}$ are the land areas for $GMPV$ and $AV$, respectively, $Y_{PV} = YY_{AV}/YY_{PV}$ is the ratio of the annual energy produced per unit module area for $AV$ to that for $GMPV$, and $\alpha = Y_{PV} \times (p/h)_{AV}/c_{M_{PV}}$. For $GMPV$, we assume $N/S$ faced modules at optimal fixed tilt for annual energy generation at $p/h = 2$. For $N/S$ faced $AV$ system having tilt identical to $GMPV$, $Y_{PV} \approx 1$ regardless of $p/h$ as long as the row-to-row shading between modules is negligible. For $E/W$ faced vertical $AV$ system, $Y_{PV} < 1$ for most locations around the world.

For a given $p/h$, we can lump the constant parameters and re-write (7) as:

$$\kappa \leq P_c' - c_L' + Y_{PV} \quad (8)$$

where $P_c'$ and $c_L'$ represent normalized crop profit and land related costs. Defining $\rho \equiv P_c' - c_L' + Y_{PV}$, respectively, we can write (8) as:

$$\kappa \leq \rho \quad (9)$$

where $\rho$ represents the overall food-energy profit for $AV$.

Eq. (9) is a key result of this paper. The criterion in (9) can be used to evaluate the technoeconomic feasibility of $AV$ relative to $GMPV$ for a range of scenarios including a variety of land costs, crop rotations and module configurations. $\kappa$ is usually >1 due to a customized foundations and elevated $AV$ mounting structure. A threshold $\rho$ ($\rho_{th}$) can be defined for the case of equality in (9) to ensure the land preservation cost is balanced by the food-energy profits. For a given module system, $\kappa$ can serve as an input parameter in (9) and $\rho_{th} = \kappa$ can be sought to ensure economic equivalence to $GMPV$. In general, the $AV$ design including suitable crops and module configurations can be optimized to maximize $\rho$ above $\rho_{th}$. For the case when the crops are pre-selected and land costs are known at the design stage, $\rho$ can be estimated as the model input and a threshold $\kappa$ ($\kappa_{th} = \rho$) can be defined to be sought by optimizing the module technology. The land preservation cost or the margin for an extra investment on $AV$ module technology relative to $GMPV$ is then equal to $(1-\rho) \times 100\%$.

Since $\kappa$ is above 1 in most practical situations, the relative economic feasibility for $AV$ does not hold for $\rho < 1$. The economic feasibility criterion in (11) implies that the cultivation of high value crops under $AV$ system is desired. In addition, an optimal choice of $p/h$, a cost-effective design for the elevated module infrastructure, and the selection of land, are important. Different module configurations and crop rotations have been evaluated in literature for $AV$ systems. The mounting structure cost can significantly vary due to different elevation, choice of materials, and the design of foundations. The practical value of $\kappa$ therefore depends upon specific economic details for a given $AV$ design. For example, $\kappa$ derived for ~5m elevated mounting installed in Germany is about 1.38 [12]. For vertical $E/W$ oriented $AV$ systems, $\kappa$ is typically lower as the minimum ground coverage of the modules allows a significantly lower elevation while still preserving most of the agricultural land. The value of $\kappa$ for vertical $AV$ from a field experiment could not be found in the literature. A recent estimate from NREL [25] however estimates 20% increase in the premium costs for vertical $AV$ as compared to GMPV. The precise value for $k$ could however vary depending upon the actual need for elevation based upon the height of the intended crops. In this work, we assume that $\kappa = 1.2$ for $E/W$ vertical bifacial $AV$, *i.e.*, 20% higher relative to $GMPV$. The qualitative findings of this paper however remain applicable for any value of $\kappa$ and are hence useful for any practical system design.

### B. Effect of Feed-in Tariff (FIT)

When the government policy allows for a higher $FIT$ for $AV$ relative to GMPV, we can add a factor of $\Delta FIT = FIT_{AV} - FIT_{PV}$ into (10) and get:

$$\left(\frac{c_{M_{AV}}}{c_{M_{PV}}}\right) \leq P_c' - c_L' + Y_{PV} + \frac{\Delta FIT}{\beta} \quad (10)$$

$$\Delta FIT \geq \beta \times (\kappa - \rho) \quad (16)$$

where $\beta = c_{M_{PV}}/(YY_{AV} \times \chi)$

The threshold feed-in tariff ($\Delta FIT_{th}$) is defined when the equality holds for (11). $\Delta FIT_{th}$ is proportional to the module technology costs per unit energy produced and depends on the difference between $k$ and $\rho$ which makes it highly dependent on the system design including modules' orientation, array density, and the land costs. $\Delta FIT$ can be used as a tool by the policy makers to support agricultural land preservation through $AV$. Moreover, $\Delta FIT$ can be made crop-specific if cultivation of some selected crops needs to be promoted at a given location.

### C. Economic condition in terms of crop profit

We can rearrange (10) to define a criterion for crop profit:

$$P_c \geq \left[c_{M_{AV}}\left(1 - \frac{Y_{PV}}{k}\right) + c_L - \frac{\Delta FIT}{YY_T \times \chi}\right] \quad (12)$$

The above criterion signifies that $AV$ crop profit needs to compensate the higher $AV$ module technology costs and the given land cost while a higher $FIT$ for $AV$ can allow the crops having relatively low value to be economic feasible. Defining $P_c \equiv Y_{PAR} \times P_{c,open}$, where $Y_{PAR}$ is the biomass yield ratio for $AV$ vs. open field, we can re-write (12) as:

$$Y_{PAR} \geq \psi \qquad (13)$$

where $\psi$ quantifies the combined effect of costs related to modules and land (minus any adjustment due to $\Delta FIT$) normalized with $P_{c,open}$. $Y_{PAR}$ represents the $AV$ crop yield relative to the open farm.

The criterion for $Y_{PAR}$ in (13) provides a simple measure for selecting crops with an appropriate market value to ensure an economic equivalence with respect to $GMPV$ at a given $FIT$, and the costs related to modules and land. Moreover, it allows to estimate an upper bound to the biomass yield loss that can be economically tolerable for $AV$. It is worth noting that $\psi$ is normalized to the crop profit which implies that the high value crops are more likely to meet the criterion in (13) as compared to the low value crops. It also confirms that the crops which have relatively small loss in biomass yield under the module shades are preferable. For a given $AV$ system, farmer and policy makers can negotiate $\Delta FIT$ which could satisfy (13) for the desired crop.

### D. Calculation of Energy and Crop Yields

The model to simulate the module energy generation and the $PAR$ available to the crops under the modules is explained in our previous papers [19, 20]. Here, we briefly describe the approach. Assuming a relatively large size module arrays and ignoring the effects at the edges of the arrays, we solve for the shading patterns in two spatial dimensions, i.e., perpendicular to the length of the arrays and the height above the ground. A view factor model which has been previously validated on field experiments [19, 20, 30] calculates the sunlight intercepted by the modules to get the temporal $PV$ yield which includes the contributions from direct beam, diffused light, and albedo (both direct and diffuse components). The shading for the direct and diffused light is calculated for the 2D planes along the vertical direction below the modules to find the $PAR$ incident on crops. Typical meteorological conditions for Khanewal, Punjab, Pakistan (30.2864° N, 71.9320° E) are used [30, 32]. The analysis and simulations are performed for $N/S$ faced 30° fixed-tilt and $E/W$ faced vertically installed bifacial systems. Different $AV$ farm schemes based on $N/S$ faced fixed tilt and $E/W$ faced vertical bifacial along with conventional Ground Mounted $PV$ ($GMPV$) systems are shown in Fig. 1 with height ($h$) and pitch ($p$) labelled.

The average income/profit for the conventional agriculture is taken from the [32] for the year 2018 for Khanewal $P_c = Y_{PAR} \times P_{c0}$. The crop yield loss due to shading is calculated based on the drop in the useful $PAR$ received by the crop across the day. The daily useful $PAR$ is calculated by taking the daily integrating of the $PAR$ received by the crop up to its light saturation point. The daily yield ratio, $Y_{PAR} = \frac{PAR_{u,AV}}{PAR_{u,open}}$, where $PAR_{u,AV}$ and $PAR_{u,open}$ are the daily useful $PAR$ for the $AV$ and the open agriculture farms, respectively. The crop yield for $AV$ is then calculated as: $Y_{c,AV} = Y_{PAR} \times Y_{c,open}$. We have previously shown that this method shows a reasonable match with $AV$ field experiments [19, 20]. It should however be noted that this approach does not consider any synergistic effects of shading on crop yield (e.g., increase in crop yield with shading for certain climate/crops under hot/arid conditions as has been reported in [8]). The approach used here can therefore be considered an upper limit for crop loss due to $AV$ shading. Nevertheless, the overall economic framework presented here is generic and can be used with any crop yield model or actual field data for crop yield.

### III. RESULTS AND DISCUSSIONS

The modeling framework is applied for two conceptual $AV$ farms: a) high value, and b) low value farms represent crop rotations that yield high and low annual profit, respectively for Khanewal (30.2864° N, 71.9320° E), Punjab, Pakistan. Each farm is studied under conventional $N/S$ faced fixed tilt and $E/W$ faced vertical bifacial module orientations. The cropping cycle and reported crop yield/revenues for Khanewal are taken into consideration while simulating the low value and high value farms. Crop rotation for the high value farm comprises of tomato, cauliflower and garlic over the year, while for the low value farm, it consists of wheat and cotton as shown in the appendix (Table I and Table II, respectively).

### A. Effect of the module density and land cost at $\Delta FIT = 0$

In this section, we explore the economic trends assuming that there is no difference in $FIT$ between $AV$ and $GMPV$. Fig. 2 shows the effect of module array density on the economic performance for both $N/S$ and E/W faced $AV$ orientations under different $M_L$. The reported values for $\kappa$ (as discussed in the section II) are shown by dotted straight lines. $\rho$ is evaluated using (9) to understand the economic implications of varying $p/h$ and $M_L$. For the high value farm, the trend for $\rho$ vs. $p/h$ changes from a negative to positive slope as $M_L$ is increased. An intermediate behavior is found at $M_L \sim 10$, for which $\rho$ first increases with $p/h$ and maximizes at $p/h \sim 3$ and then shows a downward trend. Higher $p/h$ implies more land area under $AV$ which allows an increased crop revenue. On the other hand, higher $M_L$ implies reduced land related costs which favors using more land, i.e., higher $p/h$. Fig. 2 (a) shows that the economic equivalence ($\rho_{th} = k$) with respect to $GMPV$ can be obtained for $N/S$ high value farm when $M_L$ is between 30—50 at $p/h$ of ~5.5—6. For the low value farm (Fig. 2 (c)), the trend for $\rho$ vs. $p/h$ remains with a negative slope for all $M_L$ but $\rho$ becomes lesser dependent on $p/h$ as $M_L$ is increased. The economic equivalence to $GMPV$ is however not approachable even at higher $p/h$ and $M_L$. These trends highlight the quantitative significance of $\kappa$ for the economic feasibility. Lower $\kappa$ can enhance the economic feasibility of $AV$ at lower values of $M_L$ and $p/h$. Similar trends for $\rho$ vs. $p/h$ are obtained for $E/W$ faced vertical bifacial orientation as shown in Fig. 2 (b and d). Here, the value of $\kappa$ is lower as compared to the N/S faced orientation as discussed in section II-A.

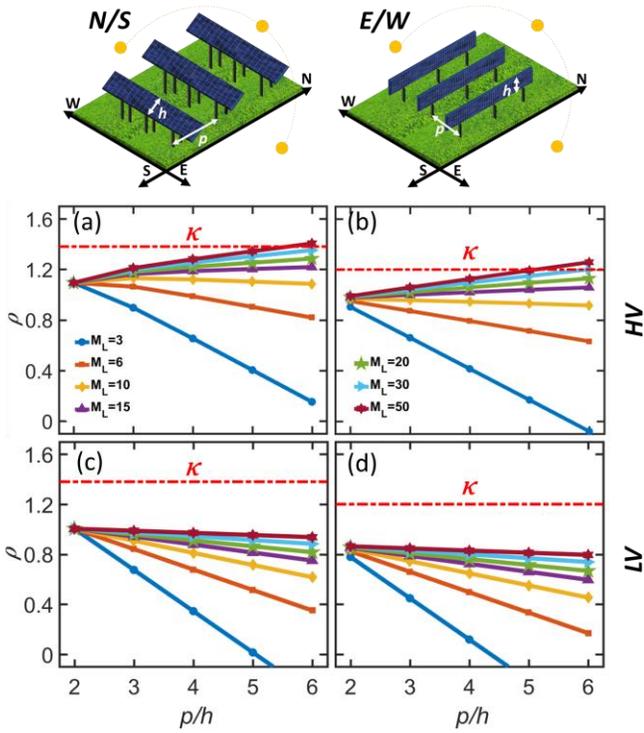

Figure 2. Normalized food-energy profit vs. $p/h$ for $N/S$ fixed tilt and $E/W$ vertical bifacial $AV$ orientations at different $M_L$ for high value and low value crops. Horizontal dotted lines ($\kappa$) represent the normalized land preservation cost for the respective $AV$ module technology.

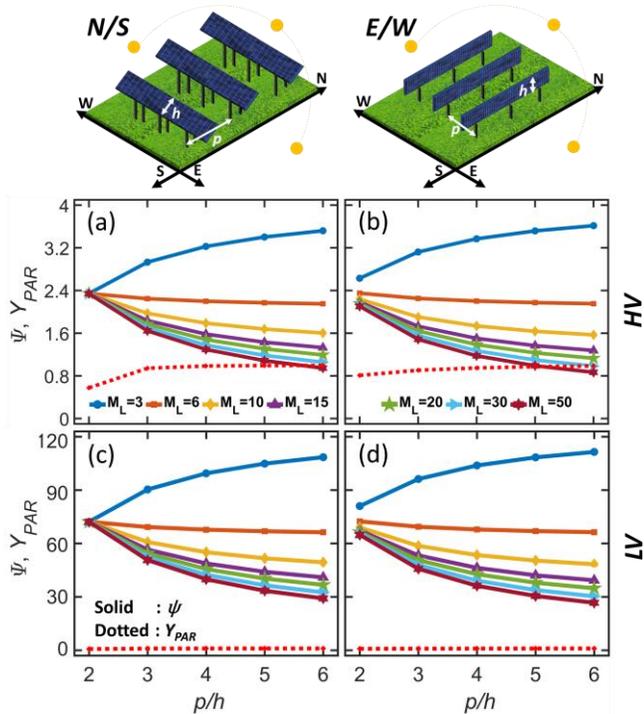

Figure 3. Normalized criteria ($\psi$) and modeled $AV$ crop yield ($Y_{PAR}$) relative to the open farm for $N/S$ fixed tilt and $E/W$ vertical bifacial $AV$ orientation at different $M_L$ for high value (top) and low value crops (bottom). The normalized criteria is satisfied for the high value crops for $p/h > 5$. For low value crops, the criteria far exceeds the crop yield for both module configurations.

Moreover, since more daily light intensity is available to crops in $E/W$ vertical as compared to the $N/S$ tilted orientation at the same $p/h$ (see Fig. A2 in the appendix), higher crop yields are predicted in $E/W$ vertical. This however has a tradeoff with the annual energy generation which is generally lower for the $E/W$ vertical in comparison with $N/S$ faced fixed tilt $AV$ systems [17]. Fig. 2 (b) shows that the economic equivalence ($\rho_{th} = k$) with respect to $GMPV$ can be obtained for $E/W$ $HV$ farm when $M_L$ is above 30 and $p/h$ is ~5—6. For $E/W$ $LV$ farm (Fig. 2 (d)), however, the economic equivalence is not approachable even at higher $p/h$ and $M_L$.

Fig. 3 explores the economic feasibility of $AV$ from the perspective of relative crop profit. $Y_{PAR}$ and $\psi$ for $HV$ and $LV$ farms as a function of $p/h$ and $M_L$ is plotted for $N/S$ and $E/W$ module orientations. For smaller $p/h$, crop yield loss due to shading can be significant depicted by smaller $Y_{PAR}$. At the same time, the normalized costs associated with land preservation have a higher impact at reduced $p/h$ since the crop revenue is limited due to smaller land area utilized under $AV$. As $p/h$ increases, $Y_{PAR}$ approaches towards 1 due to reduced shading. For high land cost ($M_L < 10$) $\psi$ increases with $p/h$, while the trend switches for higher $M_L$ as the higher crop revenue for larger $p/h$ starts playing a dominant role. For both $N/S$ and E/W orientations, the economic criterion for matching $GMPV$ is met for $HV$ at $p/h \sim 5$ and $M_L > 30$. On the other hand, $LV$ farms under both module orientations fail to meet the criterion.

Fig. 4 shows the effect of $M_L$ on $\rho$ for the $HV$ and $LV$ farms for both module orientations at $p/h = 3$. It can be observed that $\rho$ vs. $M_L$ has a sharp slope when $M_L$ is smaller (<10). For these $M_L$ (higher land related costs), $AV$ systems are difficult to be economically feasible. A saturation behavior for $\rho$ is seen when $M_L$ exceeds beyond the range of 10—20. It is worth noting that $\rho$ remains smaller than $\kappa$ for all values of $M_L$ shown in Fig. 4 implying that the economic equivalence to $GMPV$ can only be met through offering a higher $FIT$ for all the shown $AV$ designs.

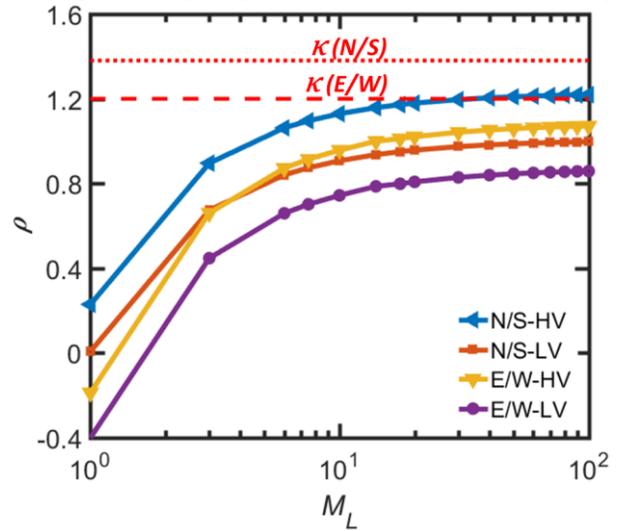

Figure 4. Effect of $M_L$ on the normalized food-energy profit for $N/S$ fixed tilt and $E/W$ vertical bifacial orientations for constant $p/h = 3$ is shown for high value and low value crops.

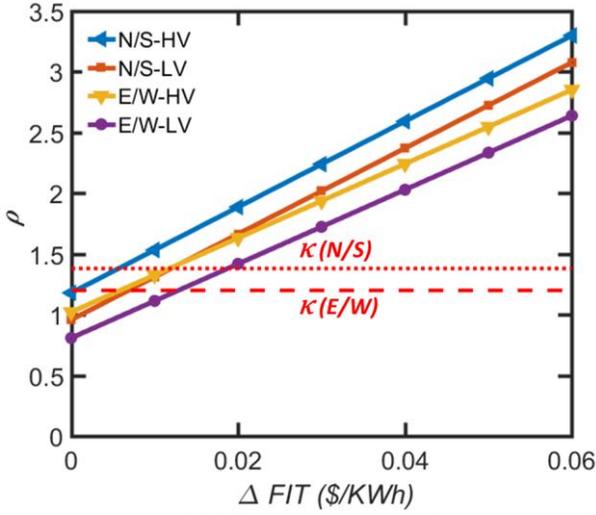

Figure 5. The impact of $\Delta FIT$ on the normalized food-energy profit for $p/h = 3$ and $M_L = 20$ is shown. Dotted horizontal lines ($\kappa$) represents normalized land preservation cost the respective module technology.

### B. Effect of Feed-in Tariff (FIT)

Until now, we have assumed an identical $FIT$ for $AV$ and $GMPV$. In practice, government policies may allow a higher $FIT$ for $AV$ to promote the land preservation. We now explore how $\Delta FIT$ can modify the trends shown in the previous section. Fig. 5 shows the effect of $\Delta FIT$ on $\rho$ for $p/h=3$ and $M_L=20$. As expected, a linear behavior between $\Delta FIT$ and $\rho$ is observed. We can compute $\Delta FIT_{th}$ for each of the module orientations and crop rotation by looking at the intersection of $\rho$ and the respective values of $\kappa$ (shown as dotted horizontal lines at for $N/S$ and $E/W$ faced orientations, respectively) taken as the inputs. $\Delta FIT_{th}$ is the lowest for the $HV$ farm in $N/S$ faced orientation closely followed by the $HV$ farm in the $E/W$ vertical orientation. For the $LV$ farm, a higher policy incentive in the form of a higher $\Delta FIT_{th}$ is required for making $AV$ system economically equivalent to $GMPV$. $\Delta FIT_{th}$ for the $HV$ farm is significantly lower as compared to that for the $LV$ farm due to a large difference in the profits obtained from crops. The difference between $\Delta FIT_{th}$ for the $N/S$ vs $E/W$ orientations for the same crop rotation is however not significant.

Fig. 6 shows the impact of $M_L$ on $\rho$ for $p/h=3$ for $N/S$ and $E/W$ $AV$ configurations. Similar to the behavior of $\rho$ vs. $M_L$ (see Fig. 5), $\Delta FIT_{th}$ shows a sharp increase when $M_L$ drops below 10 for all $AV$ configurations corresponding to the land related costs becoming significantly high. With increasing $M_L$ (lower land related costs), $\Delta FIT_{th}$ drops and later saturates as $M_L$ exceeds 20.

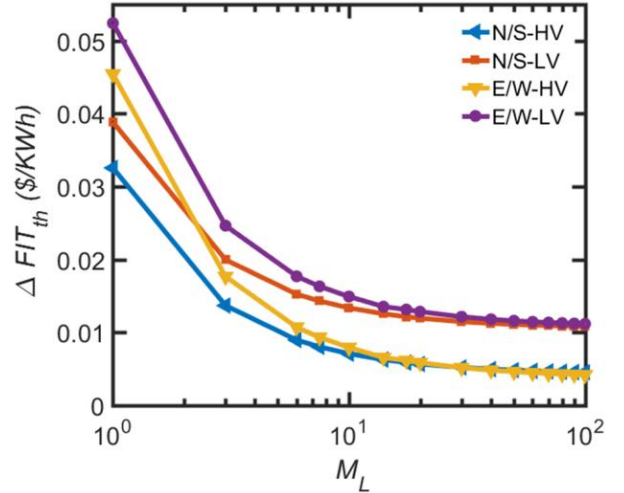

Figure 6. Effect of $M_L$ on $\Delta FIT$ for $N/S$ fixed tilt and $E/W$ vertical bifacial orientations for $AV$ is shown for high value and low value crops at $p/h = 3$.

### C. Design space

Based on the trends shown in Fig. 2 – 6, we can identify an economically feasible design space for $AV$ systems for different crop rotations, module configurations, land related cost without and with $\Delta FIT$. Fig. 7 and Fig. 8 show color contours of $\rho$ and $M_L$ respectively, as a function of $M_L$ and $p/h$ for $HV$ and $LV$ farms. The economic performance becomes equivalent or better than $GMPV$ when $\rho$ equals or exceeds the given $\kappa$ for the given module configuration. It is evident from Fig 7. that for high value (both $N/S$ and $E/W$) farms, $\rho_{th}$ is viable at higher $p/h$ and $M_L$ exceeding 48 (for $p/h = 6$) and 30 (for $p/h > 5.5$), respectively as highlighted by insets, while for low value farms, no design space for $AV$ is economically viable due to a low income from crops. It is worth noting that for the practical $M_L$ values (10—20) for $GMPV$ [12, 24, 30], $\rho_{th}$ is not attainable for the high value farms implying the need for $\Delta FIT$ for these cases if similar land related costs are assumed for $AV$.

As it is evident from Fig. 7, that there is no economically viable region for low value farm relative to $GMPV$ while high value farm may also underperform for some of the practical ranges of $M_L$, so a contribution in the form of $\Delta FIT$ is required. Fig. 8 shows the design space for various $AV$ systems with the required $\Delta FIT_{th}$ for economic equivalence with respect to $GMPV$. The practical zone for a given AV installation can be highlighted depending on the location and module configuration, and an optimal $\Delta FIT$ can be identified. A higher $\Delta FIT_{th}$ is required for lower crop income and it also varies with land related costs and $p/h$. The % increase in $\Delta FIT_{TH}$ for different module configurations, array densities, $M_L$, and crop rotations is given in Table III (see appendix). $\Delta FIT_{th}$ can serve as an important design guideline to $AV$ designers and investors. By offering variable $\Delta FIT$ for different locations and module configurations, policymakers can make it attractive to cultivate some specific set of crops at a typical location.

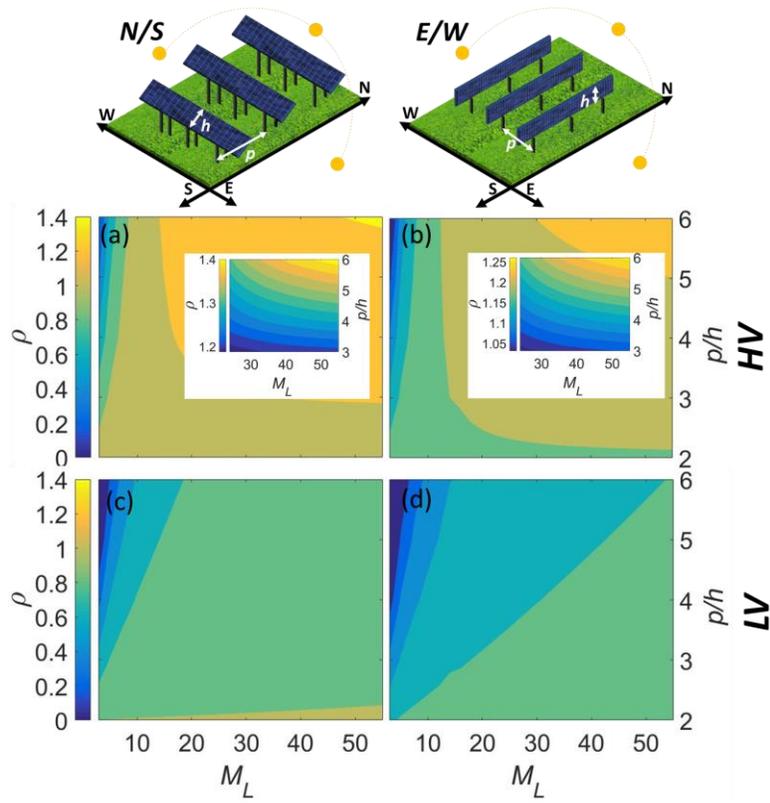

Figure 7. Normalized food-energy profit ($\rho$) as a function of $p/h$ and $M_L$ for $N/S$ fixed tilt and vertical $E/W$ bifacial orientations for high value and low value crops. Insets show a zoomed in view to identify economically feasible design space.

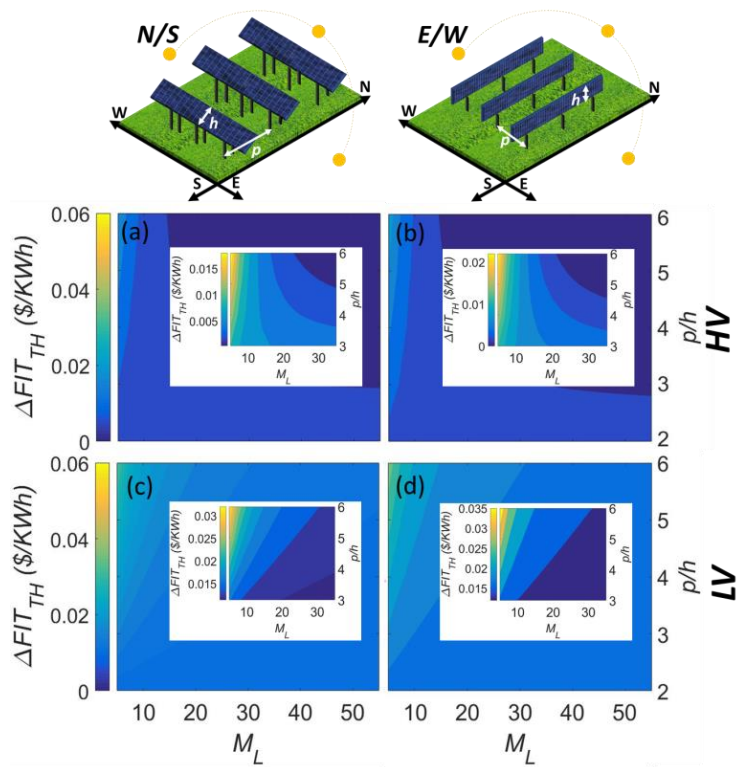

Figure 8. $\Delta FIT_{th}$ required for economic feasibility with respective to GMPV as a function of $p/h$ and $M_L$ for $N/S$ and vertical bifacial $E/W$ orientations with high value and low value crops.

*D. Limitations and future extensions*

Although the general criterion described in (9) and (13) is accurate, following assumptions are made in deriving the inequalities. **First,** the energy profit is assumed to be higher than the crop profit per unit land area which is often the case for $AV$ systems. **Second,** the response of the crop yield to the partial shading is based on the spatial-temporal light sharing ($LPF$) model described in [19]. The $LPF$ model nicely models the intra-day light fluctuations under the $AV$ shades but ignores the effects of water or nutrients stress and any shade related physiological adaptation by the crop. Nevertheless, the model provides a very convenient way to do a first order analysis especially when the purpose is to compare a variety of module configurations and other system variables. **Finally,** the model evaluates the economic feasibility assuming that the energy and food profits are owned by the same entity. This may be valid when a farmer owns the solar investment or vice versa but this does not need to be always the case. If the solar investor and the farmer are separate entities, the profits from energy and crop, and the land related costs need to be distributed among them according to the business deal. Policy interventions from the government becomes much more important under these scenarios and can strongly influence the techno-economic design space. Extensions of our model are planned to cover such scenarios and will be a part of the future research.

Future studies are planned to extend the application of this modeling framework to other crop rotations along with exploring the tracking module configurations. Tracking allows reduction of height as it would allow farm equipment movement by tilting away from the combine harvesters. The shadow-depth is also reduced and energy yield increase significantly [33]. Therefore, it can change the economics of $AV$ in important ways.

## IV. Conclusions

In this paper, we have presented a techno-economic modeling framework to assess and predict the economic performance of $AV$ systems relative to the standard ground mounted $PV$. The effects of module design configurations including array density and orientation, income from crop, technology specific and land related costs, and $FIT$ are explored. To support cropland preservation, $AV$ typically has a higher module technology cost as compared to standard $PV$ primarily due to elevated mounting and customized foundations that can potentially make it economically non-attractive for $PV$ investors. We show that it is possible to design an economically attractive $AV$ system by selecting suitable crops and module configuration for the given land costs and $FIT$. The model is applied to compare the relative economic performance of fixed tilt $N/S$ vs. $E/W$ faced vertical bifacial modules at various module densities for two selected crop rotations that represent high and low profit margin crops for southern Punjab, Pakistan. Following conclusions are made based on the modeling results:

- To offset the land preservation cost for $AV$, module arrays at a reduced (~1/3$^{rd}$) density are economically favorable with the high value crops when the land costs are relatively lower than the module costs (*i.e.*, $M_L > 20$). The crop biomass yield loss remains small under this situation because of low shading.
- For low value crops, reducing the module density is not economically desirable even when land costs are small. This implies that the standard module density can be the appropriate choice provided the crop biomass yield does not drop below an acceptable limit defined by the local policy.
- $E/W$ faced vertical module configuration can although be less productive in terms of annual energy production, its overall economic performance can match closer to the standard $N/S$ faced modules due to its lower land preservation cost. Secondary benefits (not quantified in this study) for the vertical configuration include minimum ground coverage, negligible soiling loss, and shade homogeneity for crops that can be additional merits when making the technology choice.
- For high value crops and low land costs, $AV$ can provide equivalent profitability relative to the ground mounted case without needing a higher $FIT$. When the crop profit is low, a moderate increase (~10% for the case studied here) in $FIT$ is however needed for $AV$ for economic equivalence.
- When the land costs are high and approach closer to the module costs (*i.e.*, $M_L < 10$), $AV$ economic performance shows a high sensitivity to $M_L$. This trend tends to saturate above $M_L \sim 20$.
- The design space for $AV$ to be economic equivalent to ground mounted system without a higher $FIT$ for both (E/W and N/S) module configurations needs $p/h \sim 5$, high value crop, and $M_L > 20$. As $FIT$ is increased (~10%) relative to that for the ground mounted system, low value crops, higher $p/h$, and smaller $M_L$ can be economically feasible.

In summary, this study finds that higher balance-of-system costs due to the land preservation for the cropland plays an important role in the $AV$ economics. Since $AV$ technologies are still in an early stage of development, innovations in the design of mounting including materials and structures, and the development of best practices could help reduce the land preservation cost in the future. The modeling approach in this study can remain be a valuable tool toward better understanding the economic feasibility of $AV$ as the technology develops in future. Although a simple approach is used for modeling crops in the current work, more sophisticated models and field validation can be incorporated for the crop yield changes, water use efficiency, microclimate impact on the module efficiency, changes in the operation and maintenance costs, and soiling impact will be addressed in future work.

## V. APPENDIX

### A. Effect of different orientations

Fig. A1 shows the impact of $N/S$ fixed tilt and $E/W$ vertical bifacial orientations on $Y_{PAR}$ for both high value and low value farm (for $M_L = 20$). There is almost no effect of mentioned orientations on the $Y_{PAR}$ and thus on crops for $p/h$ greater than 2. For full density ($p/h = 2$), there is significant difference in $Y_{PAR}$ of $E/W$ vertical bifacial and $N/S$ fixed tilt orientations with $Y_{PAR}$ around 30-40% higher for $E/W$ vertical bifacial orientations. This leads to higher crop yield (and thus revenue) for $E/W$ vertical bifacial orientation but does not translate into higher $\rho$ for full density ($p/h = 2$) as shown in Fig. A1. This is due to the fact that even though more light is available under $PV$ panels for $E/W$ vertical bifacial orientation (thus higher $Y_{PAR}$), but energy generated by this orientation is comparatively less than fixed tilt $AV$ system at full density ($p/h = 2$), thus resulting in lower $\rho$ even at lower panel densities for $E/W$ vertical bifacial orientation than $N/S$ fixed tilt orientation.

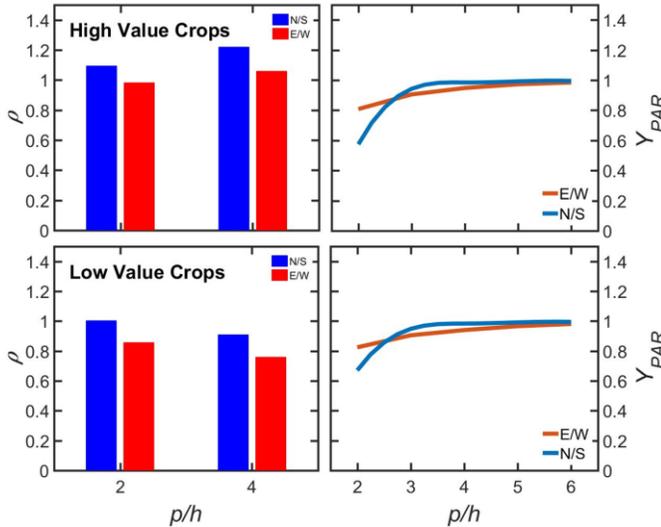

Figure A1. (left) Normalized food-energy profit for $p/h = 2, 4$ for $N/S$ fixed titl and vertical $E/W$ faced orientations. (right) Normalized crop yield $Y_{PAR}$ vs. $p/h$ for high value and low value crops.

### B. Effect of Constant Feed-in tariff

The electricity tariffs are reducing globally due to continuous improvement in $PV$ technology along with reduction in its costs [34]. In Pakistan, the feed in tariff of $PV$ is also reducing and it is becoming cheaper [35]. In recent years, $PV$ tariff in Pakistan is between 5 to 7 cents per KWh so by increasing it by 50% for $AV$ in order to meet the additional costs for $AV$, a hypothetical case is presented depicting the effect of $FIT$ for $M_L = 20$. Fig. A2 shows $\rho$ vs. $p/h$ trend in the presence of $\Delta FIT$ for $N/S$ and $E/W$ faced module orientations, respectively for the two crop rotations. Feed-in Tariff resulted in providing offset and thus shifting the trends upwards (in comparison with trends in Fig 2 and 3), thus achieving higher values of $\rho$ at lower p/h and $M_L$. Fig. A2 (a) shows that the economic equivalence ($\rho_{th} = k$) after including $\Delta FIT$ with respect to $GMPV$ can be obtained for $N/S$ high value farm for $M_L = 20$ at $p/h$ of 6. For the $N/S$ low value farm (Fig. A2 (a)), however, the economic equivalence is still not approachable implying a much higher $\Delta FIT$ is required due to low crop income. Similarly, Fig. A2 (b) shows that the economic equivalence ($\rho_{th} = k$) after including $\Delta FIT$ with respect to $GMPV$ can be obtained $E/W$ high value farm for $M_L = 20$ at $p/h$ of ~6. For the $E/W$ low value farm (Fig. A3 (b)), however, the economic equivalence is still not approachable implying a much higher $\Delta FIT$ is required due to low crop income.

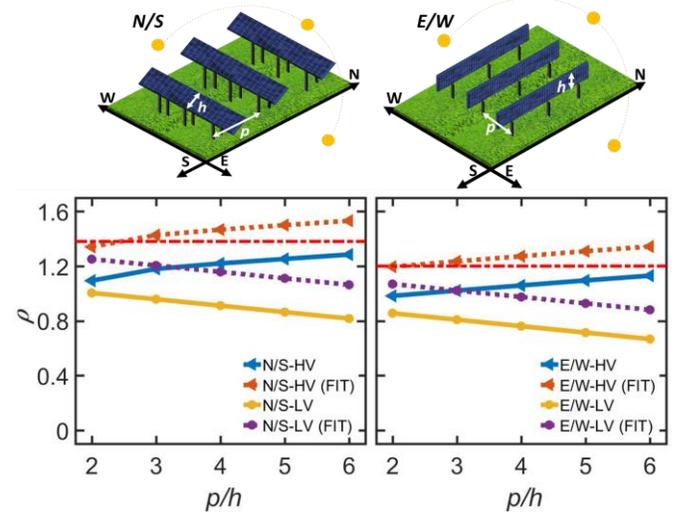

Figure A2. Normalized food-energy profit ($\rho$) at $\Delta FIT = 0.007$ \$/KWh and $M_L = 20$ as a function of $p/h$ for $N/S$ and vertical bifacial $E/W$ orientations with high value and low value crops.

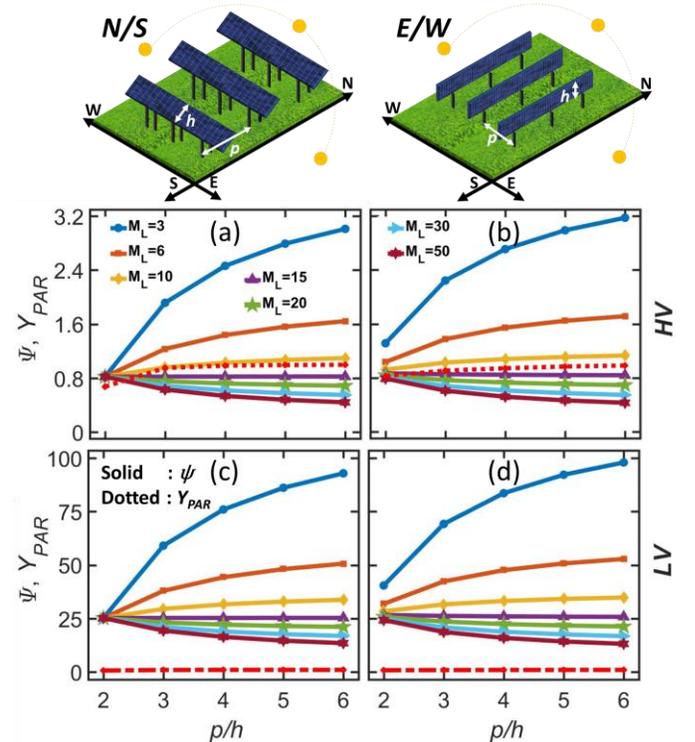

Figure A3. Normalized crop yield criteria ($\psi$) and $AV$ crop yield ($Y_{PAR}$) relative to the open farm as a function of $p/h$ for $N/S$ fixed tilt and $E/W$ vertical bifacial $AV$ orientations at different $M_L$ and with high value and low value crops.

### C. Details on Mathematical modeling:

We will start from Eq. (6) in mathematical modelling section which is given by

$$\frac{\left(M_L+\frac{p}{h}\right)_{AV}}{\left(M_L+\frac{p}{h}\right)_{PV}} \times \frac{1}{Y_{PV}} \leq \frac{P_C}{\left(M_L+\frac{p}{h}\right)_{PV} \times \left(\frac{c_L}{\chi}\right) \times A_M} + 1 \quad (i)$$

$$(M_L + p/h)_{AV} \times \frac{1}{Y_{PV}} = (M_L + p/h)_{PV} + \frac{P_C}{\left(C_L/\chi\right) \times A_M} \quad (ii)$$

$$(M_L + p/h)_{AV} = Y_{PV} \times (M_L + p/h)_{PV} + \frac{Y_{PV} \times P_C}{\left(C_L/\chi\right) \times A_M} \quad (iii)$$

$$M_{L_{AV}} - M'_{L_{PV}} = \frac{Y_{PV} \times P_C}{\left(C_L/\chi\right) \times A_M} - \left(\frac{p}{h_{AV}} - \frac{p'}{h}_{PV}\right) \quad (iv)$$

where $M_{L_{PV}}' = M_{L_{PV}} \times Y_{PV}$ & $\left(\frac{p}{h}\right)_{PV}' = \left(\frac{p}{h}\right)_{PV} \times Y_{PV}$

$$\frac{C_{M_{AV}} - C'_{M_{PV}}}{C_L} = \frac{Y_{PV} \times P_C}{\left(C_L/\chi\right) \times A_M} - \left(\frac{p}{h_{AV}} - \frac{p'}{h}_{PV}\right) \quad (v)$$

where $M_L = C_M/C_L$ [30]

$$\left(\frac{C_{M_{AV}}}{C'_{M_{PV}}}\right) = \frac{Y_{PV} \times P_C \times \chi}{C'_{M_{PV}} \times A_M} - \left(\frac{p}{h_{AV}} - \frac{p'}{h}_{PV}\right)\frac{C_L}{C'_{M_{PV}}} + 1 \quad (vi)$$

$$\left(\frac{C_{M_{AV}}}{C_{M_{PV}}}\right) = \frac{Y_{PV} \times P_C \times \chi \times p/h_{AV}}{C_{M_{PV}}} - \frac{\left(\frac{p}{h_{AV}} - \frac{p'}{h}_{PV}\right)}{M_{L_{PV}}} + Y_{PV} \quad (vii)$$

Knowing that $\frac{A_{L,PV}}{A_{L,AV}} = \frac{p}{h_{PV}}/\left(\frac{p}{h_{AV}}\right)$ where $A_{L,PV}$ is the land area occupied by GMPV and $A_{L,AV}$ is land area occupied by AV. Using $\alpha = \frac{Y_{PV} \times p/h_{AV}}{C_{M_{PV}}}$ & $\kappa = \left(\frac{C_{M_{AV}}}{C_{M_{PV}}}\right)$, we get

$$\left(\frac{C_{M_{AV}}}{C_{M_{PV}}}\right) = \alpha \times P_C \times \chi - \left(\frac{1}{Y_{PV}} - \frac{A_{L,PV}}{A_{L,AV}}\right)\alpha \cdot c_L + Y_{PV} \quad (viii)$$

## D. Crop Revenue Inputs:

Table I. Cropping cycle and net profit from Tomato, Cauliflower and Garlic (High Value Farm) for Khanewal.

| Months | Crop | Revenue ($/ha)[36] |
|---|---|---|
| Apr-Jun | Tomato | 948.81 |
| Jul-Sep | Cauliflower | 1,145.98 |
| Oct-Mar | Garlic | 7,097.54 |
| Total | | 9,192.34 |

Table II. Cropping cycle and net profit from Cotton and Wheat (Low Value Farm) for Khanewal

| Months | Crop | Revenue ($/ha)[36] |
|---|---|---|
| Apr-Sep | Cotton | 69.88 |
| Oct-Mar | Wheat | 228.43 |
| Total | | 298.31 |

## E. Feed in tariff requirement for economic equivalence:

Table III. $\Delta FIT$ (in %) required for $AV$ to achieve an economic equivalence with respect to $GMPV$ ($HV$ and $LV$ Farm) for Khanewal

| $M_L$ | $p/h$ | N/S LV | E/W LV | N/S HV | E/W HV |
|---|---|---|---|---|---|
| | | % $\Delta FIT_{TH}$ | | | |
| 10 | 2 | 15.22 | 16.79 | 11.60 | 10.85 |
| | 3 | 19.10 | 21.33 | 10.13 | 11.36 |
| | 4 | 23.03 | 25.88 | 10.53 | 11.95 |
| 15 | 2 | 15.22 | 16.36 | 11.60 | 10.43 |
| | 3 | 17.95 | 19.35 | 8.98 | 9.37 |
| | 4 | 20.72 | 22.34 | 8.22 | 8.41 |
| 20 | 2 | 15.22 | 16.15 | 11.60 | 10.22 |
| | 3 | 17.08 | 18.36 | 8.11 | 8.38 |
| | 4 | 18.99 | 20.57 | 6.49 | 6.64 |
| 30 | 2 | 15.22 | 16.15 | 11.60 | 10.00 |
| | 3 | 16.41 | 18.36 | 7.44 | 7.39 |
| | 4 | 17.64 | 20.57 | 5.14 | 4.87 |